\documentclass[twocolumn,amsmath, amssymb, jcp,showpacs, nobibnotes]{revtex4-1}
 
 \usepackage[dvips]{graphicx}
 
\usepackage[latin1]{inputenc} 
\usepackage[T1]{fontenc}

\begin{document}

\title{Potential energy curves for the interaction of Ag($5s$) and Ag($5p$) with noble gas atoms}

\author{J. Loreau, H. R. Sadeghpour, and A. Dalgarno}
\affiliation{ITAMP, Harvard-Smithsonian Center for Astrophysics, Cambridge, Massachusetts 02138, USA}
\date{\today}

\begin{abstract}
We investigate the interaction of ground and excited states of a silver atom with noble gases (NG), including helium. Born-Oppenheimer potential energy curves are calculated with quantum chemistry methods and spin-orbit effects in the excited states are included by assuming a spin-orbit splitting independent of the internuclear distance. We compare our results with experimentally available spectroscopic data, as well as with previous calculations. Because of strong spin-orbit interactions, excited Ag-NG potential energy curves cannot be fitted to Morse-like potentials. We find that the labeling of the observed vibrational levels has to be shifted by one unit.
\end{abstract}

\pacs{}

\maketitle

\section{Introduction}

There is a great deal of interest in trapping and cooling of atomic and molecular species, ostensibly for detailed manipulation of interatomic interactions, and precision spectroscopy \cite{Jones2006,Carr2009}. Cold and ultracold atomic and molecular ensembles are also employed as prototypes to simulate many-body quantum condensed-phase matter \cite{Trefzger2011}, to study processes far from equilibrium \cite{Tomadin2011}, and to create qubits for quantum logic operations \cite{DeMille2002}. Nearly all of the current focus has been on cooling and trapping of alkali metal atoms and associated molecular species, because of the availability of accessible cycling transitions for laser cooling. Alkali atoms can be treated as one-electron atoms, which makes them amenable to accurate numerical calculations of their properties. The paradigm shift to other atoms in the periodic table occurred with the advent of general-purpose magnetic and off-resonant optical trapping schemes in recent years. In magnetic trapping of atoms or molecules, a first necessary ingredient is a species with a spin projection. One such atom is silver which has been confined in a magneto-optical trap  \cite{Uhlenberg2000}, and in a buffer-gas cooled magnetic trap  \cite{Brahms2008}. It was found that in a high density He buffer gas cooled trap, Ag has a sizeable propensity to undergo three-body recombination (Ag-He-He $\rightarrow$ AgHe ($v=0,J)^*$ - He $\rightarrow$ AgHe$(0,0)$ - He) and form van der Waals (vdW) complexes \cite{Brahms2010}. This process of formation of weakly-bound molecules shows up as a loss of Ag atoms from the trap. Collisions of optically-pumped spin-polarized atoms with $^3$He have been shown to be highly efficient in transfer of spin polarization to $^3$He nuclei. Silver was shown in a recent work to be even more efficient than commonly used alkali-metal atoms for polarization transfer \cite{Tscherbul2011a}. 

The cold and ultracold molecules come in two main flavors: they are either weakly-bound highly vibrationally-excited Feshbach molecules, created by pairing ultracold atoms, or are deeply-bound molecules which can be paramagnetic, for trapping in a buffer-gas trap, or polar, for slowing in an electric field decelerator and eventually trapped. A third class of trappable molecules is the vdW molecules, which are bound solely by long-range dispersion interaction and are weakly bound. 
Among these, the interaction of Ag and other coinage metals with noble gases has been the subject of numerous experimental studies \cite{Jouvet1991,Brock1995,Brock1995b,Knight1997,Plowright2007,Plowright2008,Plowright2009,Plowright2010}. Spectroscopic studies on these complexes have focused on the molecular absorption corresponding to the strong atomic $^2P \leftarrow\, ^2S$ transition. The understanding of the bonding of such VdW complexes can be used to improve models of atom-surface interaction and study of chemical reaction dynamics \cite{Brahms2011,Balakrishnan2004,Hutson1990}, while their decay by chemical exchange, pre-dissociation and dissociation, can be controlled by external fields. 
Silver complexes with noble gas atoms can also be used for application to magnetometry \cite{Sushkov2008}, and the pressure broadening and shift of the $D_1$ line of Ag in collisions with Ar and He were recently measured \cite{Karaulanov2012}.
 
\begin{figure}[htp]
\includegraphics[width=.46\textwidth]{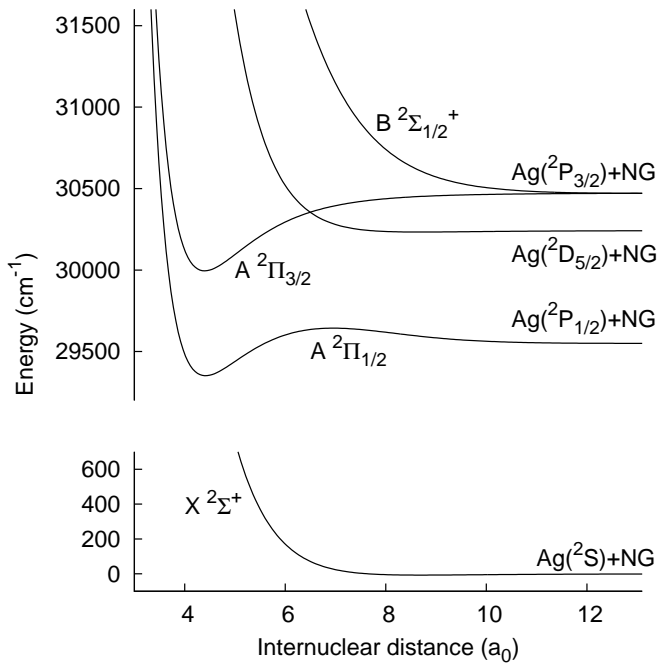}
\caption{Schematic representation of the PECs of the low-lying states of Ag-NG Van der Waals complexes.}
\label{fig_spectrum}
\end{figure}

In this work, we describe the molecular states resulting from the interaction of Ag($5s$) and Ag($5p$) with all the noble gases and focus on the Ag-Ar system in order to establish a comparison with experimental data. A schematic diagram of the potential energy curves (PECs) of the low-lying spectrum of the Ag-NG complexes, including the spin-orbit coupling, is shown in Fig. \ref{fig_spectrum}.
The ground state of these vdW complexes is the $X\ ^2\Sigma^+$ state that correlates to Ag($5s\, ^2S$) + NG($^1S$) and is attractive for all noble gases. The first excitation of the silver atom, Ag($5p\, ^2P$), gives rise to the $A\ ^2\Pi$ and the $B\ ^2\Sigma^+$ molecular states. The $B\ ^2\Sigma^+$ state is expected to be less strongly bound than the ground state as it corresponds to the interaction between the NG($^1S$) and the $p$ orbital of Ag oriented parallel to the intermolecular axis. On the other hand, in the case of the $A\ ^2\Pi$, the $p$ orbital is perpendicular to the intermolecular axis, leading to a more attractive state. Ag-He is an exception to this rule since He does not have a $p$ shell, and the absence of $p-p$ repulsion leads to an even more deeply bound $A\ ^2\Pi$ state.

The spin-orbit (SO) interaction cannot be neglected in systems involving silver. Its effect is to split the $^2P$ state of Ag into the doublet states $^2P_{1/2}$ and $^2P_{3/2}$ at 29552.1 cm$^{-1}$ and 30472.7 cm$^{-1}$, respectively, separated by 920.6 cm$^{-1}$.
In the Ag-NG complexes, the effect of the SO interaction is to mix the $A\ ^2\Pi$ and $B\ ^2\Sigma^+$ states into a $^2\Pi_{1/2}$ state (dissociating to $^2P_{1/2}$) and $^2\Pi_{3/2}$ and $^2\Sigma^+_{1/2}$ states (dissociating to $^2P_{3/2}$). 
However, the major complication in the theoretical treatment of the Ag-NG systems is the fact that the second excited state of silver corresponds to $4d^95s^2(\ ^2D_{5/2})$. This $^2D_{5/2}$ component overlaps with the $4d^{10}5p$ ($ ^2P_{3/2}$) state, rendering an accurate calculation a considerable task. Therefore, we treat the SO interaction analytically by approximating the coupling by its atomic value, and discuss for Ag-Ar the validity of this approximation.

Previous theoretical {\it ab initio} calculations of the potential energy curves of the ground state of Ag-noble gases systems include studies of Ag-He complexes \cite{Jakubek1997b,Cargnoni2008}, and Ag with He, Ne and Ar \cite{Tong2009}. Similar calculations were performed by Gardner et al. \cite{Gardner2010}, in addition to the PEC of the Ag-Kr, Ag-Xe, and Ag-Rn complexes.
The interaction potential of Ag with N$_2$ has also been recently reported \cite{Loreau2012a}, completing the study of the interaction of silver in its ground state with buffer gases.
However, except for the case of Ag-He  \cite{Jakubek1997b}, the excited states of these vdW complexes have never been investigated theoretically.
On the experimental front, two spectroscopic studies of silver-noble gases complexes exist: Jouvet {\it et al.} investigated the Ag-Ar complex using laser-induced fluorescence \cite{Jouvet1991}, while Brock and Duncan used resonance-enhanced multiphoton ionization (REMPI) technique to study Ag-Ar, Ag-Kr and Ag-Xe complexes \cite{Brock1995}. Bands were observed for the transitions $A\ ^2\Pi_{1/2} \leftarrow \ X\ ^2\Sigma^+$ and $A\ ^2\Pi_{3/2} \leftarrow \ X\ ^2\Sigma^+$ and the spectroscopic parameters of these excited states were extracted.
No $B\ \Sigma^+_{1/2} \leftarrow \ X\ ^2\Sigma^+$ bands were observed. As will be shown in section \ref{section_SO}, the excited $B\ \Sigma^+$ state, which also correlates asymptotically to Ag($5p$), is weakly bound and has an equilibrium geometry at a much larger internuclear distance, such that the transition to the ground state is not favored. 

In this work, we compute PECs for the Ag-NG complexes dissociating into Ag($5s$) + NG($^1S$) and Ag($5p$) + NG($^1S$). We  describe the computational method in Sec. \ref{section_method} and present the potential energy curves without spin-orbit in Sec. \ref{section_noSO}. The inclusion of the spin-orbit interaction as a perturbation and the resulting PECs are discussed in Sec. \ref{section_SO}, and we make a detailed comparison with experimental results in Sec. \ref{section_comp_exp}.

\section{Computational method} \label{section_method}

We described the silver atom using the aug-cc-pwCV$n$Z-PP basis set \cite{Peterson2005}, with $n$=Q,5. This basis set is based on a small core relativistic effective core potential (ECP) that replaces the $1s-3d$ core \cite{Figgen2005}, and was constructed to describe accurately the remaining 19 electrons, including core-valence correlation.
For the noble gases He, Ne and Ar, we used the aug-cc-pV$n$Z basis sets \cite{Dunning1989,Woon1993,Woon1994} with $n$=Q,5. 
Calculations involving the heavier atoms Kr, Xe and Rn were realized with a small core relativistic ECP (describing respectively the 10, 28 or 60 inner electrons of these noble gases) while the aug-cc-pV$n$Z-PP ($n$=Q,5) basis set \cite{Peterson2003b} was used in order to explicitly describe the outer-core $(n-1)spd$ shells and the $nsp$ valence shells.

The characterization of the Ag-NG interaction is improved by the inclusion of a set of ($3s3p2d2f1g$) bond functions located at midway between the two atoms. This set of functions is well suited for interactions involving noble gases \cite{Cybulski1999}. The use of bond functions removes the need for the complete basis set (CBS) extrapolation while producing results that are in good agreement with the CBS limit \cite{Tao1992}.

The $X\ ^2\Sigma^+$ and $A\ ^2\Pi$ PECs were calculated using the spin-unrestricted coupled cluster method with single, double, and perturbative triple excitations (UCCSD(T)) \cite{Knowles1993,Watts1993}, as implemented in the {\small MOLPRO} 2009.1 package \cite{Molpro2009}. The reference wave functions employed in the coupled cluster calculations were generated with the spin restricted Hartree-Fock (ROHF) method.
In these calculations, we correlated not only the valence but also the outer-core electrons. This means that for silver, the effect of the $4s^2 4p^6 4d^{10} 5s$ ($5p$) electrons was included. For He, Ne and Ar, all electrons were correlated while for Kr, Xe and Rn, the $(n-1)spd$ and $nsp$ electrons were kept active.

The PEC of the $B\ ^2\Sigma^+$ state was obtained using the configuration interaction (CI) method \cite{MolproCI1,MolproCI2}, including the Davidson correction. For these calculations, we correlated the valence and $4d^{10}$ electrons of Ag and the $nsp$ electrons of the noble gas.

The PECs were calculated on a grid of internuclear distances $R$ between 3.5 $a_0$ and 20 $a_0$. At each point, we corrected the energy using the counterpoise method in order to account for the basis set superposition error \cite{BSSE}. The energies and wave functions of the rovibrational levels  were obtained by solving the radial Schr\"odinger equation using a B-spline method \cite{Loreau2010c}. The spectroscopic constants were determined by fitting the vibrational energies to the standard form $E(v)=\omega_e(v+1/2) -\omega_ex_e(v+1/2)^2$, using a nonlinear least-squares Marquardt-Levenberg algorithm. 

If rovibrational levels close to the dissociation limit are of interest, it is necessary to know the behavior of the PECs for internuclear distances larger than 20 $a_0$ as calculated in this work. The PECs can be obtained for all internuclear distances by fitting the {\it ab initio} points to the asymptotic potential $V_{\text{as}}=-\sum_n C_n/R^n$ using the dispersion coefficients $C_6$, $C_8$ and $C_{10}$ which were previously calculated for the $X\ ^2\Sigma^+$, $A\ ^2\Pi$, and $B\ ^2\Sigma^+$ states by Zhang {\it et al.} \cite{Zhang2008}.

\section{Potential energy curves without spin-orbit interaction} \label{section_noSO}

\subsection{Ground state} 

The spectroscopic parameters of the ground state $X\ ^2\Sigma^+$ are presented in Table \ref{table_param_gs}  and compared with previous theoretical works. The results presented in Table \ref{table_param_gs} were obtained with 5Z basis sets for both Ag and the noble gases. Because bond functions were employed in this calculation, the dependence of the spectroscopic constants on the basis set is expected to be small. The use of a 5Z basis set instead of a QZ basis set modifies $D_e$ by less than 1\%, and the effect is even smaller on $R_e$. 
Interestingly, for the noble gases He, Ne and Ar, we obtain a larger value for $D_e$ with the 5Z basis set than with the QZ basis set, while the effect is reversed for the heavier gases Kr, Xe and Rn. This phenomenon can be explained by the fact that the bond functions break the hierarchy of the AV$n$Z basis sets, and it is therefore not recommended to extrapolate the results to the complete basis set limit.

The PEC of the ground state of Ag-He was already studied by various groups \cite{Cargnoni2008, Tong2009, Gardner2010, Brahms2011} using the CCSD method. As can be seen from Table \ref{table_param_gs}, our results agree quite well with previous calculations, despite using different basis sets. For Ag-Ne and Ag-Ar, the agreement between the present calculations and the results of Refs. \cite{Tong2009} and \cite{Gardner2010} is excellent, although our values for $D_e$ and $R_e$ are slightly closer to those of Ref. \cite{Gardner2010}. For Ag-Kr, -Xe, and -Rn, we find again good agreement with the values reported by Gardner et al. \cite{Gardner2010}. However, we obtain larger values for $D_e$ and smaller values for $R_e$, and the discrepancy increases with the noble gas mass. This can be explained by the fact that Gardner {\it et al.} did not correlate the inner-valence electrons of the atoms in their calculations. While this has no effect for the complexes involving He, Ne and Ar (as mentioned in Ref. \cite{Gardner2010}), this is not the case for the heavier noble gases. For Ag-Rn, the effect of core-valence interactions can be as much as 10\% of the value of $D_e$. Therefore, while freezing the core or inner-valence electrons significantly reduces computational cost, it can also lead to a dramatic underestimate of $D_e$ and an overestimate of $R_e$ for heavy complexes. 

In Table \ref{table_param_gs}, we did not include the values of $D_0$ determined experimentally \cite{Jouvet1991, Brock1995}. These values were extrapolated from transitions between the ground state of the complexes and various excited states, and are strongly isotopic- and state-dependent. For example, for Ag-Kr, Ref. \cite{Brock1995} provides values for $D_0$ between 68 and 138 cm$^{-1}$, and we believe that these values are not precise enough to allow for detailed comparison. 

\begin{figure}[htp]
\includegraphics[width=.46\textwidth]{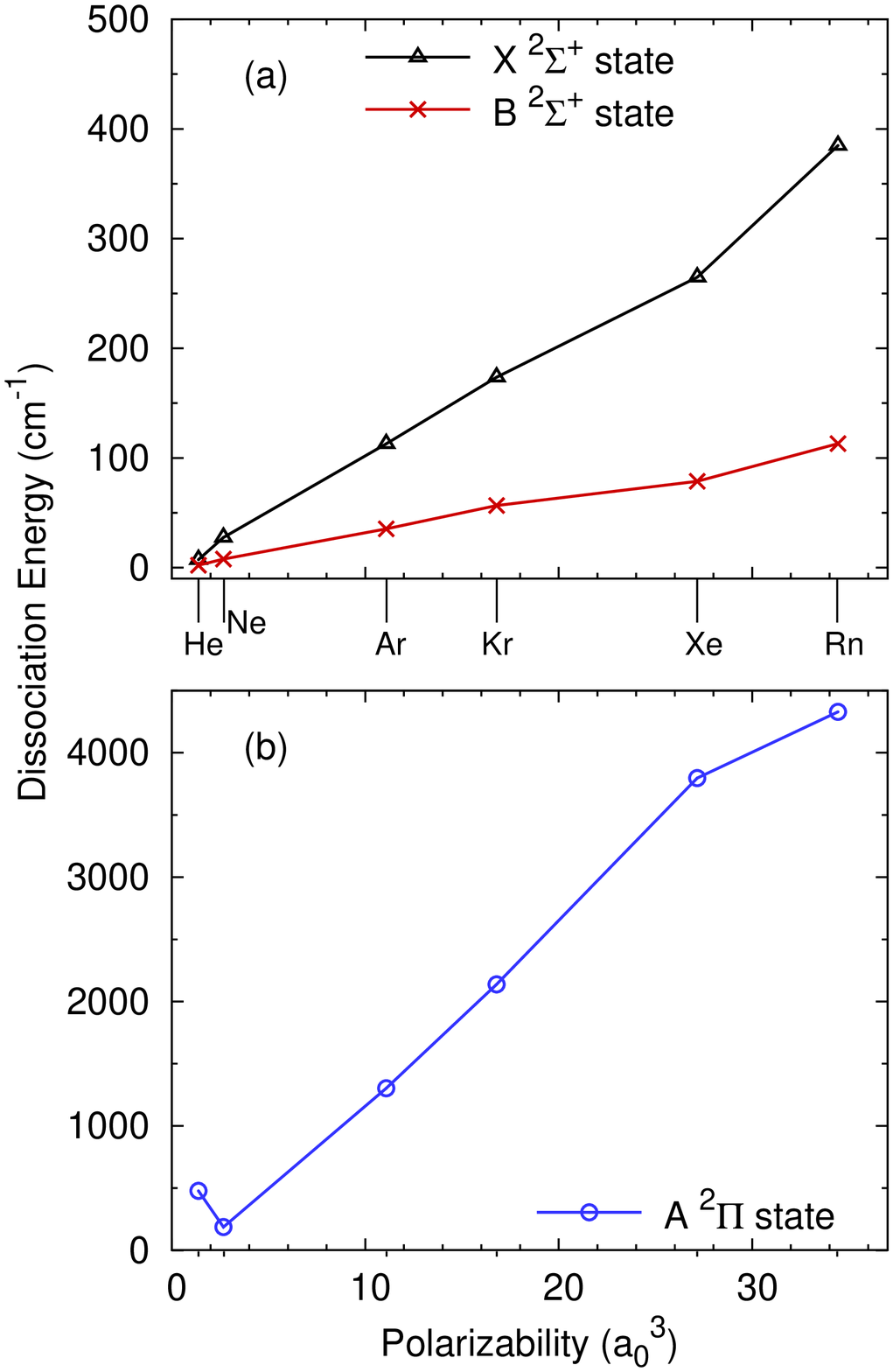}
\caption{Dissociation energy of the Ag-NG complexes as a function of the polarizability of the noble gases. (a) For the $X\ ^2\Sigma^+$ and $B\ \Sigma^+$ states; (b) for the $A\ ^2\Pi$ state. The polarizability of the noble gases increases with their mass.}
\label{fig_polarizability}
\end{figure}

The trends in the spectroscopic parameters $D_e$, $R_e$ and $\omega_e$ along the noble gas series have been discussed by Gardner {\it et al.} \cite{Gardner2010}. In particular, $D_e$ increases with NG atomic number, while $R_e$ decreases. The trend in $D_e$ is expected as the polarizability is larger for heavier noble gases, which enhances the vdW interaction. The dissociations energies are plotted in Fig. \ref{fig_polarizability}(a) as a function of NG polarizability, exhibiting a nearly linear dependence, as expected. The trend in $R_e$ is more surprising as increasing the NG mass results in an increase of the VdW radius, which in turn would be expected to lead to larger equilibrium distances. However, as extensively discussed in Ref. \cite{Gardner2010}, a combination of other factors, such as $sp$ hybridization, results in a decrease in $R_e$ along the noble gas series.

\begin{table*}[htdp]
\begin{ruledtabular}
\begin{tabular}{clllll}
Complex 	& $R_e (a_0)$	& $D_e$ 	& $D_0$ 		& $\omega_e$ 	& $\omega_ex_e$  	\\ \hline
Ag-He 	& 8.69	& 7.31		& 2.00		&			& 		\\
		& 8.67\footnotemark[1] & 7.5\footnotemark[1]	& 2.2\footnotemark[1]	& 			&		\\
		& 8.78\footnotemark[2] & 6.81\footnotemark[2]	& 					&			&		\\
		& 8.69\footnotemark[3] & 7.42\footnotemark[3]	&					& 			& 		\\
		& 8.80\footnotemark[4] & 6.80\footnotemark[4]	& 1.4\footnotemark[4]	& 			& 		\\
Ag-Ne	& 7.80	& 27.54		& 21.04		& 16.6		& 1.72	\\
		& 7.80\footnotemark[1] & 28.1\footnotemark[1]	& 21.7\footnotemark[1] & 13.5\footnotemark[1]	& 1.68\footnotemark[1] \\
		& 7.87\footnotemark[2] & 26.4\footnotemark[2]	& 				 & 13.2\footnotemark[2]	& 1.73\footnotemark[2] \\
Ag-Ar	& 7.57	& 112.93		& 102.62		& 20.0		& 0.90	\\
		& 7.53\footnotemark[1] & 113.9\footnotemark[1] & 104.2\footnotemark[1] & 19.8\footnotemark[1] & 0.88\footnotemark[1] \\
		& 7.63\footnotemark[2] & 107.2\footnotemark[2]	& 				 & 19.0\footnotemark[2]	& 0.83\footnotemark[2] \\
Ag-Kr	& 7.46	& 173.67		& 163.61		& 18.9		& 0.52	\\
		& 7.48\footnotemark[5] & 169.3\footnotemark[5] & 160.3\footnotemark[5] & 18.3\footnotemark[5] & 0.48\footnotemark[5] \\
Ag-Xe	& 7.38	& 264.81		& 254.25 		& 19.6		& 0.36	\\
		& 7.43\footnotemark[5] & 253.9\footnotemark[5] & 244.6\footnotemark[5] & 18.7\footnotemark[5] & 0.31\footnotemark[5] \\
Ag-Rn	& 7.01	& 384.80		& 372.64		& 21.7		& 0.31	\\
		& 7.13\footnotemark[5] & 355.8\footnotemark[5] & 345.8\footnotemark[5] & 20.0\footnotemark[5] & 0.26\footnotemark[5] \\
\end{tabular} 
\caption{Spectroscopic parameters of the ground state $X\ ^2\Sigma^+$ of the Ag-NG molecules. Spectroscopic parameters in cm$^{-1}$. Ag-He does not have enough vibrational levels to extract $\omega_e$ and $\omega_e x_e$.}
\label{table_param_gs}
\end{ruledtabular}
\footnotetext[1]{RCCSD(T) calculations of Ref. \cite{Gardner2010} including core-valence correlation.}
\footnotetext[2]{CCSD(T) calculations from Ref. \cite{Tong2009}.}
\footnotetext[3]{CCSDT calculations from Ref. \cite{Cargnoni2008}.}
\footnotetext[4]{RCCSD(T) calculations from Ref. \cite{Brahms2011}.}
\footnotetext[5]{RCCSD(T) calculations of Ref. \cite{Gardner2010} without core-valence correlation.}
\end{table*}

\subsection{Permanent electric dipole moments}
The vdW molecules can become polar and possess permanent electric dipole moments. In Table \ref{table_dipole_gs}, we present the vibrationally averaged dipole moments in the $X\ ^2\Sigma^+$ PEC for Ag-NG species. There is a monotonic increase of the dipole moment with the NG atom mass. Overall, the dipole moments are relatively small in the ground electronic and vibrational states, while we expect the dipole moments to be larger in the excited states.
\begin{table}[htdp]
\begin{ruledtabular}
\begin{tabular}{cc}
Complex	& Dipole moment $(D)$       \\ \hline
Ag-He	& 0.010         \\
Ag-Ne	& 0.046         \\
Ag-Ar	& 0.137         \\
Ag-Kr	& 0.201         \\
Ag-Xe	& 0.297         \\
Ag-Rn	& 0.409         \\
\end{tabular}
\caption{Permanent dipole moment of the Ag-NG species in debye.}
\label{table_dipole_gs}
\end{ruledtabular}
\end{table}

\subsection{Excited state PECs dissociating to Ag($5p\ ^2P$)}

The equilibrium distance and dissociation energy of the two excited states $A\ ^2\Pi$ and $B\ ^2\Sigma^+$, which correlate asymptotically to Ag($5p\ ^2P$) + NG, are presented in Table \ref{table_param_excited}, while their dissociation energy is plotted as a function of the noble gas polarizability in Fig. \ref{fig_polarizability}. 

The dissociation energy of the $B\ ^2\Sigma^+$ state is found to increase linearly with the noble gas polarizability, as was observed for the ground state. However, for a given complex, the value of $D_e$ is always smaller than that of the ground state. This was expected since the $B\ ^2\Sigma^+$ state corresponds to the interaction between the noble gas and the $p$ orbital of Ag oriented parallel to the intermolecular axis, which enhances repulsion. This also leads to a much larger equilibrium geometry than for the ground state. Finally, the value of $R_e$ decreases with increasing NG atomic number.

The $A\ ^2\Pi$ state corresponds to the interaction between the noble gas and the $p$ orbital of Ag oriented perpendicularly to the intermolecular axis. Therefore, this state is much more deeply bound than the $X\ ^2\Sigma^+$ or $B\ ^2\Sigma^+$ states.
The dissociation energy increases linearly from Ne to Rn, but the interaction with He results in a larger $D_e$ than with Ne (cf. Fig. \ref{fig_polarizability}(b)). This occurs because He does not have a $p$ shell, and the absence of $p-p$ repulsion leads to an even more deeply bound $A\ ^2\Pi$ state. This behavior was already observed in other systems, e.g. involving alkali metals - noble gas complexes \cite{Kerkines2002,Blank2012}.
The PEC of the $A\ ^2\Pi$ state was previously investigated for Ag-He \cite{Brahms2011,Jakubek1997b,Cargnoni2011}. Our results agree well with the calculations of Brahms {\it et al.} \cite{Brahms2011}, which were also performed using the CCSD(T) method. On the other hand, we find large discrepancies with the two other sets of calculations, performed using MP2 \cite{Jakubek1997b} and CISDT methods \cite{Cargnoni2011}, with values for $D_e$ differing by as much as 50\%.

\begin{table}[h!]
\begin{ruledtabular}
\begin{tabular}{ccll}
State & Complex	& $R_e (a_0)$	& $D_e$ (cm$^{-1}$)		\\ \hline
$A\ ^2\Pi$
& Ag-He 	& 4.40	& 477.8		\\
& 		& 4.42\footnotemark[1] & 463.6\footnotemark[1]	\\
& 		& 4.76\footnotemark[2] & 272.1\footnotemark[2]	\\
& 		& 5.16\footnotemark[3] & 349.9\footnotemark[3]	\\
& Ag-Ne	& 5.54	& 187.4		\\
& Ag-Ar	& 5.18	& 1302.6		\\
& Ag-Kr	& 5.19	& 2138.6		\\
& Ag-Xe	& 5.23	& 3796.8		\\
& Ag-Rn	& 5.34	& 4328.3		\\
\hline
$B ^2\Sigma^+$
& Ag-He 	& 13.80	& 2.30			\\
& 		& 14.91\footnotemark[1] & 0.95\footnotemark[1]	\\
& Ag-Ne	& 12.35	& 7.89			\\
& Ag-Ar	& 11.86	& 35.40			\\
& Ag-Kr	& 11.64	& 56.61			\\
& Ag-Xe	& 11.77	& 78.72	 		\\
& Ag-Rn	& 11.34	& 113.04			\\
\end{tabular} 
\caption{Spectroscopic parameters of the excited $A\ ^2\Pi$ and $B\ ^2\Sigma^+$ states of the Ag-NG molecules.}
\label{table_param_excited}
\end{ruledtabular}
\footnotetext[1]{RCCSD(T) calculation from Ref. \cite{Brahms2011}.}
\footnotetext[2]{CISDT calculations from Ref. \cite{Cargnoni2011}.}
\footnotetext[2]{MP2 calculations from Ref. \cite{Jakubek1997b}.}
\end{table}

\section{Potential energy curves with spin-orbit interaction} \label{section_SO}

As previously mentioned, the spin-orbit interaction cannot be neglected in the $5p$ shell of the silver atom as the splitting between the $^2P_{1/2}$ and the $^2P_{3/2}$ states is $\Delta=920.6$ cm$^{-1}$. 
The most accurate description of the molecular states arising from Ag($5p\ ^2P_{1/2}$) and Ag($5p\ ^2P_{3/2}$) would be realized by performing a CASSCF + MRCI calculation, including the spin-orbit interaction. However, this approach is computationally demanding as the coupling with the $\Sigma^+_{1/2}$, $\Pi_{3/2}$ and $\Delta_{5/2}$ states arising from Ag($4d^95s^2\ ^2D_{5/2}$) must also be taken into account. Therefore, we instead assume that the spin-orbit Hamiltonian is given by the atomic interaction, $H_{\textrm{so}}=\xi {\bf l\cdot s}$, where the spin-orbit parameter $\xi$ is $R$-independent. This approximation has been previously used with success to describe systems in which the spin-orbit interaction makes a full MRCI calculation intractable \cite{Jakubek1997b,Cohen1974,Plowright2010,Cargnoni2011}.
While this approximation will clearly fail at small distances, where the excited state PECs are mostly repulsive, we will test the merits of this approximation by comparing molecular parameters and vibrational energies with other calculations, when available, and observations. 
In order to obtain the spin-orbit coupled PECs, it is necessary to evaluate the matrix elements of the spin-orbit Hamiltonian. $H_{\textrm{so}}$ is diagonal in the quantum number $\Omega=\Lambda+\Sigma$ and its matrix elements can therefore be easily computed in the $\vert LSJ\Omega \rangle$ representation. However, the {\it ab initio} calculations are performed in the $\vert LS\Lambda\Sigma\rangle$ (spin-uncoupled) representation, so that it is necessary to transform the matrix elements of $H_{\textrm{so}}$ in this representation. 
From the $^2\Sigma^+$ and the $^2\Pi$ states dissociating into Ag($5p$), we can form the $^2\Pi_{3/2}$ state that correlates asymptotically to $^2P_{3/2}$ and has projection $\vert\Omega\vert=3/2$ onto the internuclear axis, and the $^2\Pi_{1/2}$ and $^2\Sigma^+_{1/2}$ states, which dissociate respectively into $^2P_{1/2}$ and $^2P_{3/2}$ and correspond to the $\vert\Omega\vert=1/2$ projection. 

\begin{figure}[htp]
\includegraphics[angle=-90,width=.46\textwidth]{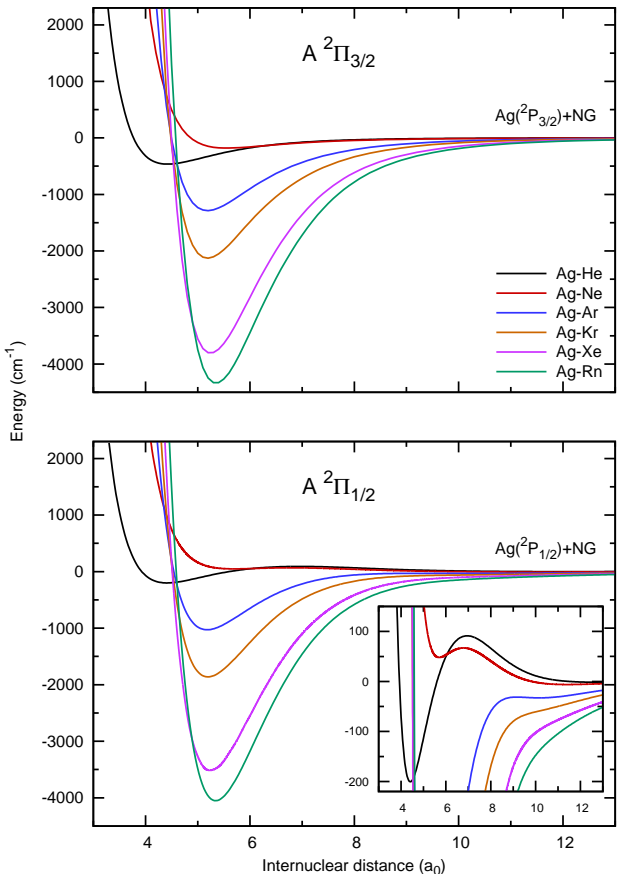}
\caption{Potential energy curves of the $A\ ^2\Pi_{3/2}$ (top panel) and $A\ ^2\Pi_{1/2}$ (bottom panel) states of the Ag($5p$)-NG complexes. The inset shows the double well structure that appear for He and Ne as a consequence of the spin-orbit interaction.}
\label{fig_AgNG_pi_SO}
\end{figure}

\begin{figure}[htp]
\includegraphics[angle=-90,width=.46\textwidth]{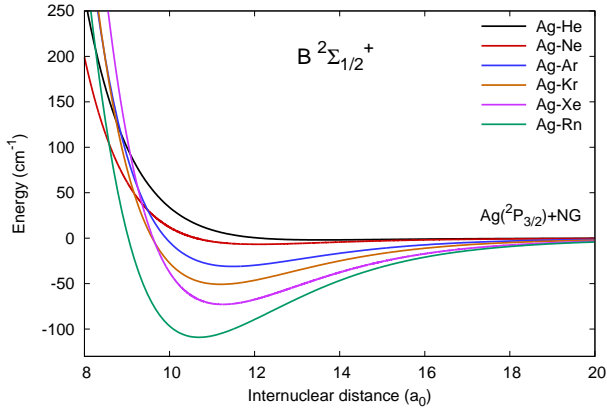}
\caption{Potential energy curves of the $B\ ^2\Sigma^+_{1/2}$ state of the Ag($5p$)-NG complexes.}
\label{fig_AgNG_sig1_2}
\end{figure}

The total Hamiltonian in the $\vert LS\Lambda\Sigma\rangle$ representation is block-diagonal in $\Omega$ and has the following form:
\begin{subequations}
\label{SOmatrix}
\begin{eqnarray}
&& \vert\Omega\vert=1/2: \quad 
\left(\begin{array}{cc}
U_{\Pi}(R) - \frac{1}{2}\xi 	& \frac{\sqrt{2}}{2}\xi \\
\frac{\sqrt{2}}{2}\xi		& U_{\Sigma}(R)
\end{array}\right) \\
&& \vert\Omega\vert=3/2: \quad
\left(\begin{array}{c}
U_{\Pi}(R) + \frac{1}{2}\xi
\end{array}\right)
\end{eqnarray}
\end{subequations}
where $U_{\Pi}(R)$ and $U_{\Sigma}(R)$ denote the PECs of the $^2\Pi$ and $^2\Sigma^+$ states, respectively.
The spin-orbit parameter is equal to two thirds of the atomic splitting, $\xi = 2\Delta/3 =613.7$ cm$^{-1}$. The diagonalization of (\ref{SOmatrix}) yields the SO-coupled potentials. 
The resulting PECs are presented in Fig. \ref{fig_AgNG_pi_SO} and Fig. \ref{fig_AgNG_sig1_2}, and the spectroscopic parameters are given in Table \ref{table_param_SO}. 

The $^2\Pi_{3/2}$ state, due to symmetry, is not affected by the SO interaction. The values of $R_e$ and $D_e$ are therefore identical to those discussed in Sec. \ref{section_noSO}. We do not observe a general trend in the behavior of $\omega_e$ for this state, while $\omega_e x_e$ decreases with increasing NG atomic number. 

The $^2\Pi_{1/2}$ PEC arises from the mixing of $p\pi$ and $p\sigma$ orbitals by the SO interaction. Since in the $^2\Pi$ molecular symmetry, the $p\pi$ orbital is perpendicular to the internuclear axis, the mixing in of the $p\sigma$ orbital results in repulsion when the Ag and NG atoms interact at short internuclear distances. 
For Ag-He and Ag-Ne, orbital repulsion is large enough to overcome the attractive character, leading to a short-range barrier and a double well structure illustrated in the inset of Fig.~ \ref{fig_AgNG_pi_SO}. This behavior was already observed in alkali-noble gas PECs \cite{Blank2012}.
The $^2\Pi_{1/2}$ state of Ag-He presents two minima separated by a barrier peaking at $R=6.95 a_0$ with a maximum energy of 91.2 cm$^{-1}$. The first minimum is located at $R=4.41 a_0$, which is almost the same value as in the $^2\Pi_{3/2}$ state. However, the dissociation energy is less than half that of the $^2\Pi_{3/2}$ state (200.6 cm$^{-1}$ compared to 477.8 cm$^{-1}$). Moreover, the $^2\Pi_{1/2}$ state only supports one bound vibrational level, whereas the $^2\Pi_{3/2}$ state supports 6 bound levels. The second minimum is located at large internuclear distance, $R_e=13.02 a_0$, and is due to the interaction with the $B\ ^2\Sigma^+$ state. This well has a dissociation energy of $D_e=1.63$ cm$^{-1}$ but does not support any vibrational levels. 
The PEC of the $^2\Pi_{1/2}$ state of Ag-Ne also presents a barrier, located at $R=6.78 a_0$ with a height of 67 cm$^{-1}$. The first minimum of the potential is situated at $R=5.70 a_0$ but has positive energy. The second minimum is located at $R_e=11.41 a_0$ with a dissociation energy $D_e=6.14$ cm$^{-1}$, which is due to the $^2\Sigma^+$ state. It supports 2 vibrational levels.
For the heavier Ag-NG species, the attractive character is strong enough so that the PEC of the $^2\Pi_{1/2}$ state is purely attractive. We find that the equilibrium distance is identical to that of the $^2\Pi_{3/2}$ state, but that the dissociation energy is systematically smaller. 
The main effect of the spin-orbit interaction is to dramatically modify the intermediate- and long-range part of the PEC, as shown in the inset of Fig.~ \ref{fig_AgNG_pi_SO}. Because of this, the PEC of the $^2\Pi_{1/2}$ state cannot be fitted to a Morse-like potential and the spectroscopic parameters $\omega_e$ and $\omega_e x_e$ presented in Table~\ref{table_param_SO} do not provide an accurate representation of the potential. This is true in particular for the high vibrational levels which lie close to the dissociation limit.

Finally, the effect of the spin-orbit interaction on the PEC of the $^2\Sigma^+_{1/2}$ state is to reduce the value of the equilibrium distance and the dissociation energy by a few percent compared to the PEC of the $B\ ^2\Sigma^+$ state without spin-orbit. For Ag-He and Ag-Ne, the potential does not support enough vibrational levels (0 and 2, respectively) to extract spectroscopic parameters.

\begin{table*}[htdp]	
\begin{ruledtabular}
\begin{tabular}{cccccccc}
State		& Complex 	& $R_e (a_0)$	& $D_e$ (cm$^{-1}$)	& $D_0$ (cm$^{-1}$) 		& $\omega_e$ (cm$^{-1}$) 	& $\omega_ex_e$ (cm$^{-1}$)  	\\ \hline
$^2\Pi_{1/2}$
& Ag-He 	& 4.41, 13.02	& 200.6, 1.63	& 119.2	&		& 		\\
&		& 5.22\footnotemark[1]		& 89.2\footnotemark[1]		&		&		&		\\
& Ag-Ne	& 11.41		& 6.14		& 3.89	& 		&	 	\\
& Ag-Ar	& 5.18		& 1028.3		& 977.0	& 106.2	& 2.76	\\
& Ag-Kr	& 5.19		& 1861.8		& 1810.1	& 103.4	& 1.44	\\
& Ag-Xe	& 5.23		& 3514.5		& 3455.1	& 118.1	& 1.00	\\
& Ag-Rn	& 5.34		& 4050.8		& 3993.7	& 111.0	& 0.77	\\ 
\hline
$^2\Sigma^+_{1/2}$
& Ag-He 	& 13.62		& 1.61		& 		&		& 		\\
& Ag-Ne 	& 12.10		& 6.70		& 4.20	& 		& 		\\
& Ag-Ar 	& 11.49		& 31.1		& 26.19	& 8.05	& 0.51	\\
& Ag-Kr 	& 11.20		& 50.8		& 45.13	& 7.89	& 0.31	\\
& Ag-Xe 	& 11.25		& 72.8		& 65.82	& 8.27	& 0.23	\\
& Ag-Rn 	& 10.69		& 109.1		& 100.93	& 8.83	& 0.18	\\
\hline
$^2\Pi_{3/2}$
& Ag-He 	& 4.40		& 477.8		& 392.6	& 174.3	& 16.1	\\
& Ag-Ne 	& 5.54		& 187.4		& 163.4	& 47.0	& 3.07	\\
& Ag-Ar 	& 5.18		& 1302.6		& 1250.9	& 100.6	& 2.00	\\
& Ag-Kr 	& 5.19		& 2138.6		& 2087.1	& 100.6	& 1.21	\\
& Ag-Xe 	& 5.23		& 3796.8		& 3738.0	& 115.5	& 0.89	\\
& Ag-Rn 	& 5.34		& 4328.3		& 4272.0	& 109.3	& 0.70	\\
\end{tabular} 
\caption{Spectroscopic parameters of the excited $^2\Sigma^+_{1/2}$, $^2\Pi_{1/2}$, and $^2\Pi_{3/2}$ states of the Ag-NG molecules for natural abundances. For Ag-He in the $^2\Pi_{1/2}$ state, the values of $R_e$ and $D_e$ for the two potential wells are given (see text). The $^2\Pi_{1/2}$ and $^2\Sigma^+_{1/2}$ states of Ag-He and Ag-Ne do not support enough vibrational states to extract $\omega_e$ and $\omega_e x_e$.}
\label{table_param_SO}
\end{ruledtabular}
\footnotetext[1]{MP2 calculations of Ref. \cite{Jakubek1997b}.}
\end{table*}

\section{Comparison with the experiment} \label{section_comp_exp}

The experimentally determined  spectroscopic parameters \cite{Brock1995,Jouvet1991} of the $^2\Pi_{1/2}$ and $^2\Pi_{3/2}$ states for specific isotopes of the Ag-Ar, Ag-Kr, and Ag-Xe complexes are presented in Table \ref{comp_expt_AgAr} -- Table \ref{comp_expt_AgXe} and compared with our calculations. These parameters are the Morse vibrational terms, $\omega_e$ and $\omega_e x_e$ (the energy of the levels is given by $E(v)=\omega_e(v+1/2) -\omega_ex_e(v+1/2)^2$), the dissociation energy 
$D_0$ ({\it i.e.}, the binding energy of the $v=0$ state), obtained by a Birge-Sponer extrapolation, and the transition frequency $\nu_{00}$ (between the $v=0$ level of either the $^2\Pi_{1/2}$ or the $^2\Pi_{3/2}$ state and the $v=0$ level of the ground state), also extrapolated using Birge-Sponer analysis. 
The dissociation energy of the ground state can also be extrapolated using the same method.

We observe that the agreement between theory and experiment for the $^2\Pi_{3/2}$ PEC is good, more so considering that the $^2\Pi_{3/2}$ PEC should interact with the PEC dissociating into Ag($^2D_{5/2}$) + NG through an avoided crossing, as illustrated schematically in Fig. \ref{fig_spectrum}. On the other hand, for the $^2\Pi_{1/2}$ state, which is expected to be the most strongly affected by the spin-orbit interaction, the parameters do not show the same level of agreement. We note that the largest discrepancy between theory and experiment occurs for the Ag-Kr system, for which the experimental data are subject to large errors \cite{Brock1995}.

\begin{table}[htdp] 
\begin{ruledtabular}
\begin{tabular}{cccc|ccc}
				&	\multicolumn{3}{c}{$^2\Pi_{1/2}$}		& \multicolumn{3}{c}{$^2\Pi_{3/2}$}  		\\ 
Param.			&	This work			& Exp. 1 & Exp. 2 		& This work				& Exp. 1 & Exp. 2	\\ \hline
$\omega_e$		& 106.3		& 109.2	& 112.9	& 100.6	& 100.3	& 100.2	\\
$\omega_e x_e$	& 2.77		& 2.83	& 3.33 	& 2.00	& 2.04	& 2.01	\\ 
$D_0$			& 977		& 999	& 903	& 1251	& 1184	& 1199	\\
$\nu_{00}$		& 28677		&	 	& 28714	& 29324	& 		& 29325	\\
\end{tabular} \caption{Comparison of the spectroscopic parameters calculated in this work with the experimental values for $^{107}$Ag-$^{40}$Ar. All parameters are in units of cm$^{-1}$.
Exp. 1 and Exp. 2 corresponds to the experimental values determined in Refs. \cite{Jouvet1991} and \cite{Brock1995}, respectively.}
\label{comp_expt_AgAr}
\end{ruledtabular}
\end{table}

\begin{table}[htdp] 
\begin{ruledtabular}
\begin{tabular}{ccc|cc}
				& \multicolumn{2}{c}{$^2\Pi_{1/2}$}		& \multicolumn{2}{c}{$^2\Pi_{3/2}$}	\\ 
Param.			&	This work			& Exp.		& This work				& Exp.	\\ \hline
$\omega_e$		& 103.8			& 121.6		& 101.1			& 108.1	 \\
$\omega_e x_e$	& 1.45			& 1.58		& 1.23			& 1.26	\\ 
$D_0$			& 1809.8			& 2286		& 2086.8			& 2267	\\
$\nu_{00}$		& 27905			& 27404		& 28549			& 28274		\\
\end{tabular} \caption{Comparison of the spectroscopic parameters calculated in this work with the experimental values from Ref. \cite{Brock1995} for $^{107}$Ag-$^{83}$Kr. All parameters are in units of cm$^{-1}$.}
\label{comp_expt_AgKr}
\end{ruledtabular}
\end{table}

\begin{table}[htdp] 
\begin{ruledtabular}
\begin{tabular}{ccc|cc}
				& \multicolumn{2}{c}{$^2\Pi_{1/2}$}		& \multicolumn{2}{c}{$^2\Pi_{3/2}$}	\\ 
Param.			&	This work			& Exp.		& This work				& Exp.	\\ \hline
$\omega_e$		& 118.9			& 123.8		& 116.2			& 115.8	 \\
$\omega_e x_e$	& 1.01			& 1.01		& 0.90			& 0.91	\\ 
$D_0$			& 3455			& 3728		& 3738			& 3630	\\
$\nu_{00}$		& 26352			& 26100		& 26989			& 27021	\\
\end{tabular} \caption{Comparison of the spectroscopic parameters calculated in this work with the experimental values from Ref. \cite{Brock1995} for $^{107}$Ag-$^{129}$Xe. All parameters are in units of cm$^{-1}$.}
\label{comp_expt_AgXe}
\end{ruledtabular}
\end{table}

The source of the discrepancy can be traced to the different behavior of SO-coupled $^2\Pi_{1/2}$ and $^2\Pi_{3/2}$ PECs, as discussed in the previous section. 
While the $^2\Pi_{3/2}$ PEC can be represented by a Morse-like potential, this is not the case for the $^2\Pi_{1/2}$ state, as was already mentioned in Ref. \cite{Brock1995}. Therefore, the values of the spectroscopic parameters $\omega_e$ and $\omega_e x_e$ cannot be expected to reflect correctly the properties of the potential, especially close to the dissociation limit.
Furthermore, the parameters $D_0$ and $\nu_{00}$ are extracted using a Birge-Sponer extrapolation, which is not expected to be particularly accurate as these parameters depend more on the low vibrational levels, while the experimental data terminate on the low end at $v=7$. 

In order to establish a more comprehensive comparison between theory and experiment, we focus for the remainder of this work on the Ag-Ar complex. Rather than comparing with the experimentally-determined spectroscopic parameters, we find it more instructive to study directly the vibrational energies of the observed transitions, for which accurate values have been reported \cite{Brock1995}.

Table \ref{table_vib_spacing_agar_pi1_2} contains the energy between successive vibrational levels in the $^2\Pi_{1/2}$ state. These energies are determined from the reported transition frequencies between vibrational levels ($v$ and $v^\prime$) in the  $^2\Pi_{1/2}$  PEC, and the ground electronic and vibrational state. At first glance, the agreement is not satisfactory. However, Brock and Duncan  \cite{Brock1995} state that there might be an error of $\pm1$ in their assignment of the vibrational levels, which is based on the isotopic shift. If we assume that the experimental levels are shifted by one unit ({\it i.e.}, the level $v=7$ is now the level $v=8$, and so forth), the agreement between theory and experiment is excellent: the average discrepancy is about 1 cm$^{-1}$. We believe that this is not a coincidence and that the assignment of the experimental vibrational levels should be shifted by unity.

\begin{table}[htdp]
\begin{ruledtabular}
\begin{tabular}{ccc}
$v\rightarrow v^\prime$	& Exp. \cite{Brock1995}	& Theory 	\\ \hline
$8 \rightarrow 7$		& 57.9	& 63.5	\\
$9 \rightarrow 8$		& 53.0	& 58.5	\\
$10 \rightarrow 9$		& 47.3	& 53.4	\\
$11 \rightarrow 10$		& 41.3	& 48.1	\\
$12 \rightarrow 11$		& 33.5	& 42.6	\\
$13 \rightarrow 12$		& 27.3	& 36.6	\\
$14 \rightarrow 13$		& 18.4	& 29.6	\\
$15 \rightarrow 14$		& 11.0	& 19.7	\\
\end{tabular} \caption{Comparison of the vibrational spacing in the $^2\Pi_{1/2}$ state of $^{107}$Ag-$^{40}$Ar with the experimental values of Ref. \cite{Brock1995}. The spacings are obtained from $(v - v^{\prime\prime})  - (v^\prime - v^{\prime\prime})= v - v^{\prime}$, where $v$, $v^{\prime}$ are the vibrational levels in the  $^2\Pi_{1/2}$  PEC, and $v^{\prime\prime}$ is a label for vibrational levels in the $X\ ^2\Sigma_{1/2}^+$ PEC.}
\label{table_vib_spacing_agar_pi1_2}
\end{ruledtabular}
\end{table}

\begin{table}[htdp]
\begin{ruledtabular}
\begin{tabular}{cccc}
$v$	& Theory		& Theory, shifted	& Exp.	\\ \hline
0	& 647.2		& 550.1		& 		\\
1	& 648.5		& 556.1		& 		\\
2	& 650.0		& 562.3		& 		\\
3	& 651.7		& 568.8		& 		\\
4	& 653.6		& 575.6		& 		\\
5	& 655.9		& 582.6		& 		\\
6	& 658.5		& 590.0		& 		\\
7	& 661.5		& 597.9		& 595.2	\\
8	& 665.1		& 606.4		& 605.2	\\
9	& 669.3		& 615.8		& 616.5	\\
10	& 674.4		& 626.2		& 628.4	\\
11	& 680.6		& 638.0		& 641.6	\\
12	& 688.3		& 651.9		& 657.9	\\
13	& 698.0		& 668.9		& 676.7	\\
14	& 711.0		& 692.9		& 708.9	\\
15	& 731.1		& 728.7		& 730.5	\\
16	& 763.1		& 758.4		&		\\
17	& 789.2		& 784.5		&		\\
18	& 811.8		& 807.5		&		\\
19	& 831.5		& 827.4		&		\\
\end{tabular} 
\caption{Vibrational dependence of the spin-orbit splitting $\delta_{\text{so}}(v)$ in the $^2\Pi$ state, $\delta_{\text{so}}(v) = E_v (^2\Pi_{3/2})-E_v(^2\Pi_{1/2})$. The second column contains the results of our calculations, while the results of the third column are obtained by shifting the numbering of the vibrational levels in the $^2\Pi_{1/2}$ by one unit of $v$. The last column contains the experimental results of Ref. \cite{Brock1995}.}
\label{table_vib_SOspacing}
\end{ruledtabular}
\end{table}

We also compared the vibrational dependence of the spin-orbit splitting in the $^2\Pi$ state, {\it i.e.} $\delta_{\text{so}}(v) = E_v (^2\Pi_{3/2})-E_v(^2\Pi_{1/2})$. These results are presented in Table \ref{table_vib_SOspacing}. We find once again that our results agree well with the observed level separation only if the vibrational levels are shifted by unity in the $^2\Pi_{1/2}$ state, but not in the $^2\Pi_{3/2}$ state. 
We can also see that $\delta_{\text{so}}(v)$ increases monotonically with $v$. This conclusion contradicts that presented in Ref. \cite{Brock1995}, where it is stated that $\delta_{\text{so}}(v)$ presents a minimum for $v=7$, but should increase for lower $v$. However, this conclusion was based on the assumption that the potential can be fitted to a Morse potential, which is not the case. 
We obtain a value of $\delta_{\text{so}}(0)=647$ cm$^{-1}$ for the $0-0$ band, larger than the expected value of $2/3\Delta = 614$ cm$^{-1}$.

Despite the large number of electrons in these systems and the approximation of an $R$-independent spin-orbit splitting, we obtain an excellent agreement with the experiment for both the $^2\Pi_{1/2}$ and the $^2\Pi_{3/2}$ states, provided that the assignment of the vibrational levels of the $^2\Pi_{1/2}$ state is shifted by one unit.

\section{Conclusions}
We calculated the Born-Oppenheimer potential energy curves for the ground and excited state interaction of silver and noble gas atoms and compared the spectroscopic parameters with available theoretical and observed data. We discussed the effect of the spin-orbit interaction using a simple model and showed that spectroscopic parameters extracted from a Morse-like potential are not accurate for the case of the PEC of the $^2\Pi_{1/2}$ state.
We obtained good agreement with experimental data for the $^2\Pi_{3/2}$ state and we showed for the case of Ag-Ar that the observed vibrational assignment in the excited $^2\Pi_{1/2}$ PEC should be shifted by one unit.

\acknowledgments
This work was supported by the U.S. Department of Energy and by an NSF grant to ITAMP at Harvard University and the Smithsonian Astrophysical Observatory.

\end{document}